# Discovery of the high thermoelectric performance in low-cost $Cu_8SiS_xSe_{6-x}$ argyrodites

Taras Parashchuk[1,&], Oleksandr Cherniushok[1,&], Raul Cardoso-Gil[3], Janusz Tobola[2], Yuri Grin[3], and Krzysztof T. Wojciechowski[1*]

[1]Thermoelectric Research Laboratory, Department of Inorganic Chemistry, Faculty of Materials Science and Ceramics, AGH University of Science and Technology, Mickiewicza Ave. 30, 30-059 Krakow, Poland
[2]Max-Planck-Institut für Chemische Physik fester Stoffe, Nöthnitzer Str. 40, 01187 Dresden, Germany
[3]Faculty of Physics and Applied Computer Science, AGH University of Science and Technology, 30-059 Krakow, Poland

**ABSTRACT**

Cu-based argyrodites have gained much attention as a new class of thermoelectric materials for energy harvesting. However, the phase transition occurring in these materials and low energy conversion performance limited their broad application in thermoelectric converters. In this work, we disclose a newly discovered highly efficient $Cu_8SiS_xSe_{6-x}$ argyrodite with stabilized high-symmetry cubic phase at above 282 K opening the practical potential of this material for the mid-temperature region applications. The temperature range broadening of the high-symmetry phase existence was possible due to the successful substitution of Se with S in $Cu_8SiS_xSe_{6-x}$, which enhances the configurational entropy. The developed argyrodites show excellent thermoelectric performance thanks to the increased density of states effective mass and ultralow lattice thermal conductivity. Further tuning of the carrier concentration through the Cu-deviation improves the thermoelectric performance significantly. The dimensionless thermoelectric figure of merit *ZT* and estimated energy conversion efficiency η for $Cu_{7.95}SiS_3Se_3$ achieve outstanding values of the 1.45 and 13 %, respectively, offering this argyrodite as a low-cost and Te-free alternative for the thermoelectric energy conversion applications.

**Keywords:** argyrodites; crystal structure; thermoelectric properties; phase transitions; electronic transport.

## 1. INTRODUCTION

Global energy production using fossil fuels wastes more than 60 % of its total energy in the form of heat emitted [1]. This process is also accompanied by the emission of toxic pollution and greenhouse gases [2]. Thermoelectric (TE) technologies allow waste heat recovery through the Seebeck effect [3], enabling electrical energy production in an eco-friendly manner [4]. This type of energy conversion does not require movable equipment, hence it is a noiseless way to produce energy with a long period of maintenance without service [5].

The energy conversion performance of the TE materials is determined by the dimensionless figure of merit $ZT = \alpha^2 T/\rho(\kappa_e+\kappa_L)$ [6], where $\alpha$, $\rho$, $T$, $\kappa_e$, and $\kappa_L$ are the Seebeck coefficient, electrical resistivity, absolute temperature, electronic and lattice thermal conductivity, respectively. The increase in the $ZT$ is a complicated task due to interdependence between the $\alpha$, $\rho$, and $\kappa_e$ facilitating the development of many strategies for decoupling the electrical properties and reduction of the only independent parameter $\kappa_L$ [7–9]. The other parameter representing the energy conversion performance, i.e. TE quality factor $B$ (which is proportional to the $\mu_w/\kappa_L$, where $\mu_w$ is the weighted mobility of charge carriers) also underlines the necessity to optimize independently the electronic and thermal transport [10,11]. Band degeneracy [12], low band effective mass [13], low deformation potential coefficient [14], in-gap states [15], and resonant states [16] approaches have been approved to be effective methods for the enhancement of electronic transport. The strengthening of the phonon scattering in thermoelectric materials is usually realized through grain boundaries [17], dislocations [18,19], nano-precipitates [20], and point defects [7].

The aforementioned strategies have shown their great potential in enhancing $ZT$ parameters in many well-established materials [21,22] and also provide a guide for discovering new thermoelectrics [23,24]. Particular attention here should be dedicated to the recent "phonon liquid electron crystal" (PLEC) concept [25]. The materials that follow this concept behave as the "phonon liquid" due to the cation migration in a crystal lattice, while superior electronic transport is realized for them through the rigid covalent framework [26].

Argyrodites are the prominent class of PLEC chalcogenides crystallized in tetrahedrally close-packed structures with the general chemical formula $A^{m+}_{(12-n)/m}B^{n+}Q^{2-}_6$ ($A^{m+}$ = Li$^+$, Cu$^+$, Ag$^+$; $B^{n+}$ = Ga$^{3+}$, Si$^{4+}$, Ge$^{4+}$, Sn$^{4+}$, P$^{5+}$, As$^{5+}$; $Q^{2-}$ = S, Se, Te) [27]. Due to the actual requirements of the new energy technologies, the large interest nowadays is dedicated to low-cost and Te-free argyrodites. One of the important problems of these materials is the polymorphic phase transition, which usually exists in most superionic conductors above room temperature. Below the phase transition, these materials usually crystallize in the low-symmetry monoclinic, orthorhombic, or cubic modifications, while the high-symmetry hexagonal or cubic modification is favorable at higher temperatures.

Except for physical property stability reasons, the high-symmetry crystal structure is desired for the TE materials due to the particularities of the band structure [28,29]. The high symmetry of the crystal lattice is reflected in the density of states (DOS) effective mass $m^*$ ($m^* = N_v^{2/3} m_b^*$, where $N_v$ and $m_b^*$ are the number of degenerated carrier pockets and the DOS effective mass at each pocket, respectively) through the increase in the number of band extrema in the corresponding point of the Brillouin zone [30]. The typical increase in $m_b^*$ causes the band flattening, which results in the decrease of carrier mobility $\mu$ according to the Bardeen-Shockley theory [31]. In turn, for high-symmetry crystals, the increase of the DOS effective mass $m^*$ is realized due to the higher band degeneracy $N_v$, hence it does not provoke the degradation of the carrier mobility $\mu$. Consequently, the increase in the number of degenerated carrier pockets enhances not only the Seebeck coefficient $\alpha$ but the power factor $\alpha^2/\rho$ [13,32].

The promising way to enhance the crystal symmetry is lying through the increase of the configurational entropy $\Delta S$ of the system. Particularly, the increase in $\Delta S$ for the low-symmetry multicomponent materials facilitates the crystal structure disorder resulting in higher symmetry of the system [29]. In such a case, the polymorphic phase transition temperatures decrease following the dependence between Gibbs free energy $\Delta G$, enthalpy $\Delta H$, and entropy $\Delta S$ ($\Delta G = \Delta H - T\Delta S$) [33]. The effect of the decrease of the polymorphic phase transition temperature from the low-symmetry modification to the high-symmetry one in argyrodites was observed in the literature [30,34] as well as in our previous work for the case of $Cu_7P(S_xSe_{1-x})_6$ argyrodites [32].

In this work, we for the first time explore the thermoelectric performance of the $Cu_8SiS_xSe_{6-x}$ argyrodites. The applied S/Se substitution in $Cu_8SiS_xSe_{6-x}$ enables a sufficient increase in the configuration entropy effectively shifting the polymorphic phase transition below the room temperature and providing the existence of the high-symmetry modification over the operating temperature range. Further adjustment of the carrier concentration was realized through the Cu-deficiency in $Cu_{8-\delta}SiS_3Se_3$. Resulting of the enhanced electronic transport and ultralow lattice thermal conductivity, the offered here $Cu_{8-\delta}SiS_xSe_{6-x}$ argyrodites are among the best compromises of the low price and high energy conversion performance ever reported for thermoelectric materials.

## 2. EXPERIMENTAL SECTION

### 2.1. DFT calculations

…

### 2.2. Preparation

Two series of samples with the nominal compositions of $Cu_8SiS_xSe_{6-x}$ ($x = 0 \div 6$) and $Cu_{8-\delta}SiS_3Se_3$ ($\delta = 0.025 \div 0.2$) were prepared. The synthesis of all samples was carried out in graphite-coated (to avoid possible reaction with $SiO_2$) quartz ampoules. The ampoules were subjected to rigorous purification, which included washing in $1HNO_3:3HCl$ concentrated acids mixture and frequent cleaning with distilled water and isopropanol, and finally dried. High-purity elements Cu (Alfa Aesar, 99.99%), Si (Alfa Aesar, 99.999%), S (Alfa Aesar, 99.999%), and Se (Alfa Aesar, 99.999%) were used for the synthesis. The total mass of each sample was 8 g. The stoichiometric amounts of elements were sealed in evacuated quartz ampoules (with an inner diameter of 18 mm and 2 mm wall thickness) and heated to 1473 K with a temperature rate of 2 K/min, kept at this temperature for 5 h, and cooled down with furnace to room temperature. The resultant dark grey and brittle ingots were ground to powder, cold-pressed, and annealed for 170 h at 873 K in quartz ampoules under vacuum. After the homogenization annealing, the ampules were cooled down with a furnace to room temperature.

The obtained samples were crushed into fine powders by hand milling in an agate mortar and then densified by Spark Plasma Sintering (SPS) technique at 1023 K for 60 min in 10 mm diameter graphite mold under axial compressive stress of 60 MPa in an argon atmosphere. The heating and cooling rates during the SPS procedure were 70 and 15 K/min, respectively. The density of the SPS-compacted cylinders with a diameter of 10 mm and length of 12 mm was more than 96 % of the theoretical density determined by the Archimedes method. From the initial cylinders, disks with a thickness of 2 mm were cut and polished for thermal diffusivity measurements. The remaining parts of the samples were cut into bars 10×10×2 mm and used for transport property measurements.

**2.3. Characterization**

Phase identification was performed with a BRUKER D8 Advance X-ray diffractometer using Cu $K\alpha$ radiation ($\lambda$ = 1.5418 Å, $\Delta 2\Theta$ = 0.005 °, $2\Theta$ range 10 – 100 °) with Bragg-Brentano geometry. The lattice parameters were determined by the least-squares refinement using the WinCSD program package [35].

Thermal analysis of the investigated samples was performed via a Differential Scanning Calorimetry equipment (Netzsch DSC 404 C and DSC 404 F3 Pegasus) using a sample mass of 28-32 mg in sealed $SiO_2$ ampoules and Al crucibles covered with a lid. The measurements were carried out with a heating rate of 10 K/min under argon or helium flow.

For microstructural analysis using optical microscopy (OM) and scanning electron microscopy (SEM), samples were embedded in conductive resin (PolyFast, Struers, Denmark), and subsequently polished, finally using 0.1 µm diamond powder in a slurry. The sample homogeneity and local chemical composition were analyzed using scanning electron microscopy (JEOL JSM-6460LV Scanning Electron Microscope and CAMECA SX100 Electron Microprobe) equipped with standard-based wavelength dispersive X-ray spectroscopy (WDS).

The Seebeck coefficient $\alpha$ and electrical resistivity $\rho$ were measured by commercial apparatus NETZSCH SBA 458 Nemesis. Measurements were made in argon flow over the temperature range of 298-773 K. Thermal diffusivity $\alpha_D$ was measured by the NETZSCH LFA 457 equipment, and the specific heat capacity $C_p$ was estimated using the Dulong-Petit limit. The samples were firstly spray-coated with a thin layer of graphite to minimize errors from the emissivity of the material and laser beam reflection caused by a shiny pellet surface. Thermal conductivity was calculated using the equation $\kappa = \rho C_p \alpha_D$, where $\rho$ is the density obtained by the Archimedes method at the cylinders from SPS. The uncertainty of the Seebeck coefficient and electrical resistivity measurements was 7 % and 5 % respectively; the uncertainty of the thermal diffusivity measurements was 3 %. The combined uncertainty for the determination of the thermoelectric figure of merit *ZT* was around 20 % [36]. It should be mentioned that the measurements of the Seebeck coefficient $\alpha$, electrical resistivity $\rho$, and thermal diffusivity $\alpha_D$ were done along the pressing direction and using the same sample for each type of material. The Hall effect was investigated by applying the four-probe method in constant electric and magnetic fields (*H* = 0.9 T) and current through a sample of 100 mA. The uncertainty of Hall measurements was around 10 %.

## 3. RESULTS AND DISCUSSION

### 3.1. Results of the DFT calculations

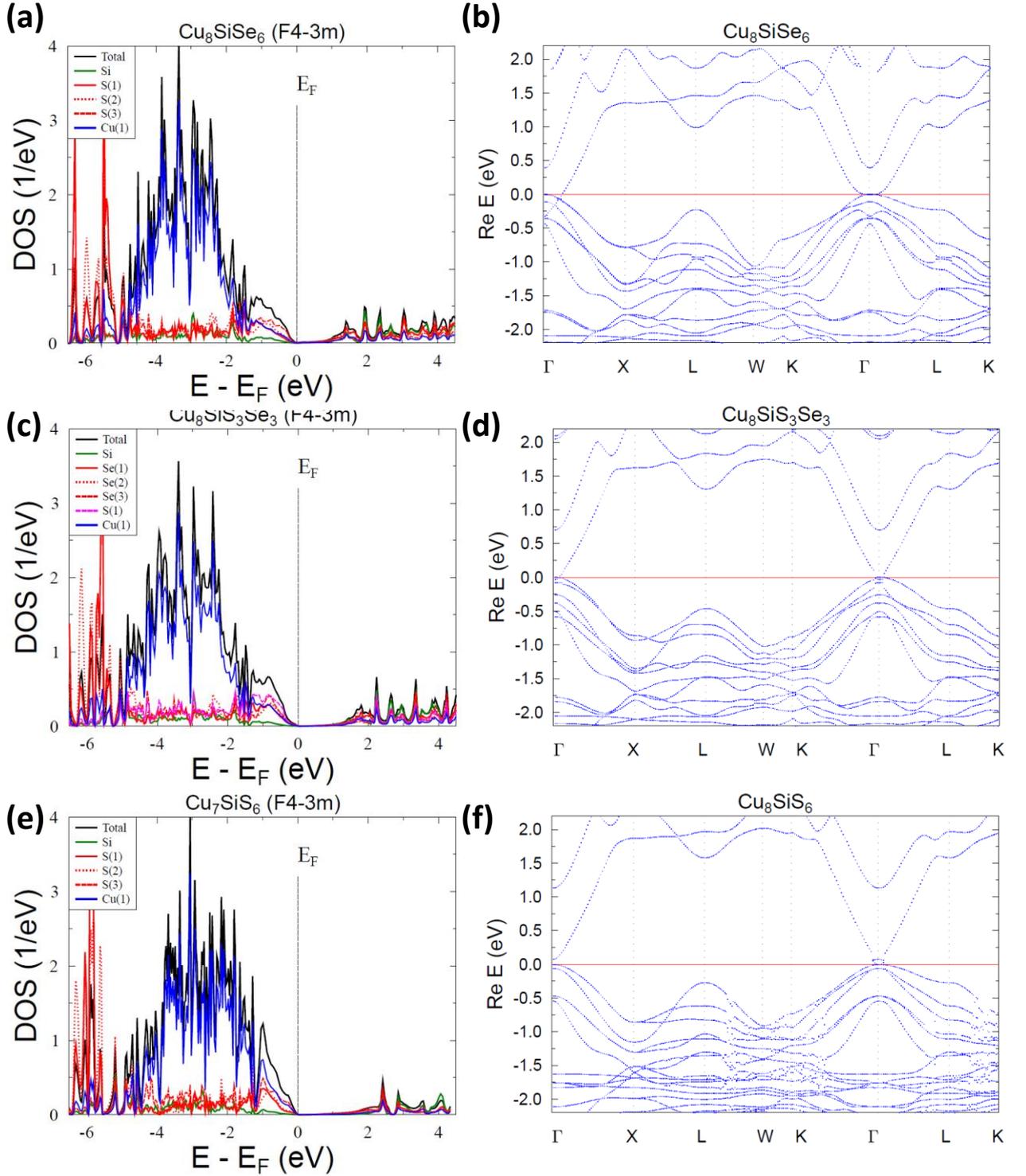

**Figure 1.** Calculated electronic structure (left panels) and band structure (right panels) of high-temperature γ-polymorphic modification (space group F-43m) for $Cu_8SiS_xSe_{6-x}$ alloys: (a,b) $x = 0$, (c,d) $x = 3$, (e,f) $x = 6$.

## 3.2. Crystal structure and phase analyses

The phase composition of the investigated $Cu_8SiS_xSe_{6-x}$ ($0 \leq x \leq 6$) samples was characterized by powder X-ray diffraction. While most of the observed XRD reflections for all samples can be indexed in orthorhombic $Pmn2_1$ or cubic $F$-43$m$ space groups (i.e. α or γ polymorphic argyrodite modifications, respectively), the minor presence of $Cu_2(S,Se)$ binaries can still be detected for the majority of $Cu_8SiS_xSe_{6-x}$ samples. Taking into account that the investigated samples were annealed for 170 h at 873 K, the presence of binary phases may indicate that studied argyrodites crystallize with a deviation from the stoichiometry.

Powder XRD patterns recorded at room temperature for $Cu_8SiS_xSe_{6-x}$ samples after synthesis are shown in Figure 2a. With the increase of the sulfur content, the positions of the reflections monotonically shift to higher $2\theta$ values indicating a decrease in the lattice parameters due to the partial substitution of Se by smaller S atoms. The $Cu_8SiSe_6$ and $Cu_8SiS_6$ powder XRD patterns were indexed in the orthorhombic $Pmn2_1$ space group, which corresponds to the low-temperature α-modification [27]. These results agreed well with the data, which is shown by Kuhs *et al*., that the ordered α-$Cu_8SiSe_6$ ($Pmn2_1$) undergoes a superionic phase transition at 323 K into a disordered high-temperature γ-$Cu_8SiSe_6$ ($F$-43$m$) polymorph, while in the case of $Cu_8SiS_6$, a superionic phase transition takes place at 336 K [27]. Hereby, both compounds should exist in the low-temperature α-modification at room temperature XRD experiment, and the phase transition is expected with the increase in temperature.

The reflections of only high-temperature γ-polymorph (space group $F$-43$m$) were observed at the powder XRD patterns for $Cu_8SiS_xSe_{6-x}$ samples with $x = 1$, 2, and 3, while the samples with $x = 4$ and 5 show the presence of both α and γ polymorphs (Figure 2a). Consequently, the samples with $x = 1$, 2, and 3 have the largest interest for the thermoelectric applications, as they show the high-temperature argyrodite phase already at room temperature, and the phase transition is not expected for them with temperature increase up to the melting point (Figure S1).

The lattice parameters of $Cu_8SiS_xSe_{6-x}$ samples obtained by the least-squares refinement of the XRD reflection position in the range of $10° \leq 2\Theta \leq 100°$ are shown in Figure 2b. The smaller atomic radius of S compared to Se leads to the decrease of lattice parameters indicating the successful substitution of Se by S. The linear decrease of the lattice parameters over the whole compositional range indicates that selenium can be fully replaced with sulfur in $Cu_8SiS_xSe_{6-x}$.

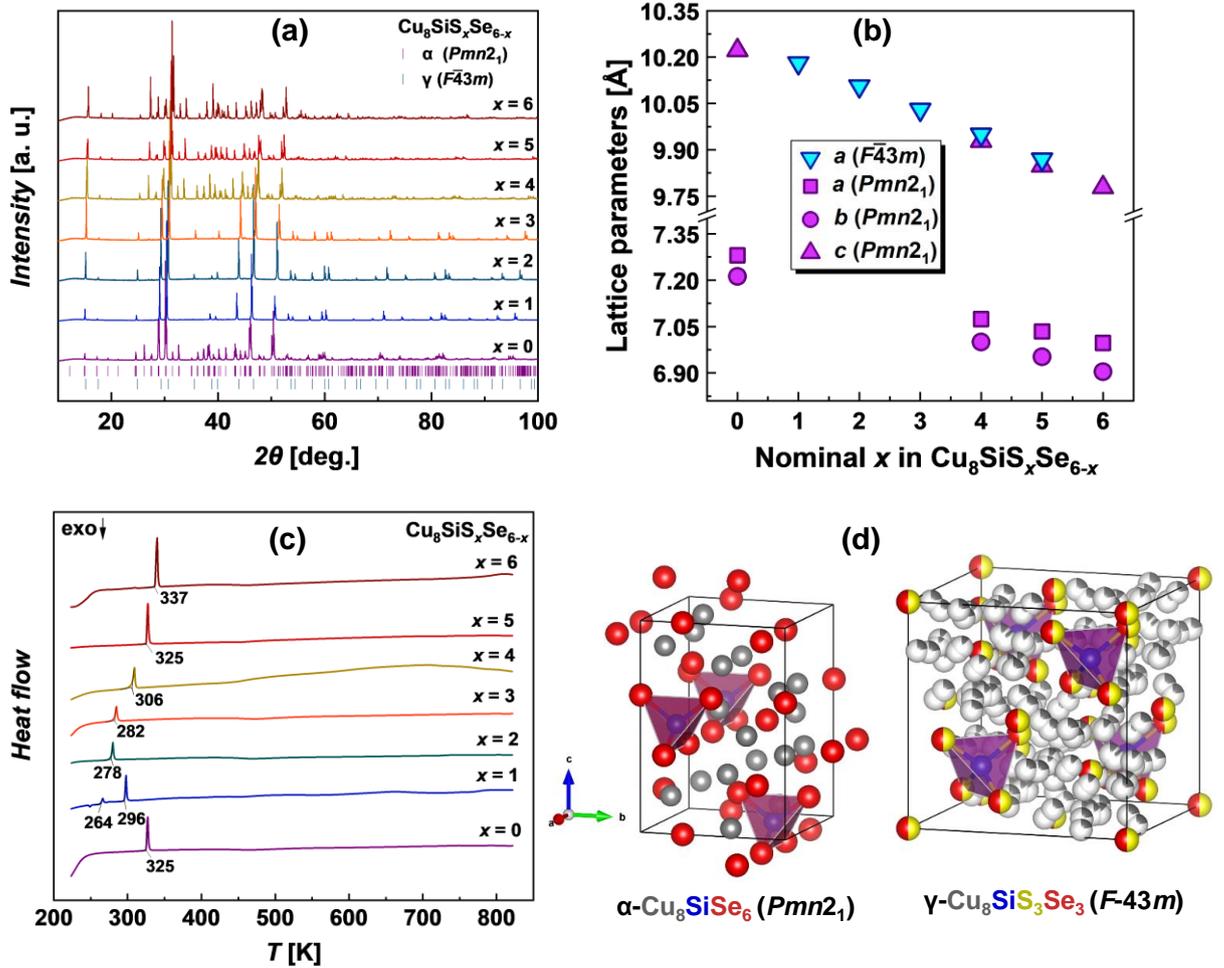

**Figure 2.** (a) Powder X-ray diffraction patterns, (b) lattice parameters, and (c) DSC curves for the $Cu_8SiS_xSe_{6-x}$ materials; (d) representation of the crystal structure for orthorhombic α-$Cu_8SiSe_6$ (space group $Pmn2_1$) and cubic γ-$Cu_8SiS_3Se_3$ (space group $F\bar{4}3m$) polymorphic modifications.

The thermal behavior of the investigated materials was studied with the help of differential scanning calorimetry (DSC). Particularly, to provide a high sensitivity of the signal and accurately analyze the phase transition region for the investigated argyrodites, the DSC measurements were made using the Al crucibles over the temperature range of 200-823 K (Figure 2b). In turn, to estimate the melting temperatures, we also performed the DSC measurements in sealed quartz ampules from room temperature to above the melting temperatures up to 1523 K (Figure S1). The ternary $Cu_8SiS_6$ and $Cu_8SiSe_6$ argyrodite compounds melt congruently at 1445 K and 1350 K, respectively (Figure S1). The reported melting temperatures for $Cu_8SiS_6$ and $Cu_8SiSe_6$ are 1459 K and 1380 K, respectively, which are close to the achieved results [37]. The recorded DSC curve of the sample with $x = 0$ shows an endothermic peak at 325 K, which corresponds to polymorphic phase transition α-$Cu_8SiSe_6$ ↔ γ-$Cu_8SiSe_6$. This result is in good agreement with the previously reported phase transition temperature for $Cu_8SiSe_6$ [27]. The temperature of the polymorphic phase transition decreases with S substitution showing 278 K for the sample with $x = 2$, then again increases to 337 K for the sample with $x = 6$. Performed DSC measurements (Figure 2b) additionally prove that for samples with $x = 1$, 2, and 3, the high-temperature γ-polymorphic ($F$-43$m$) modification is stable even at room temperature, as it was suggested by the XRD analysis (Figures 2a and d). Interestingly, the existence of the α phase reflections for $Cu_8SiS_xSe_{6-x}$ samples with $x = 0$, 5, and 6 was observed at the XRD patterns after the melting of materials during DSC measurements, while only γ-phase phase reflections were presented for samples with $x = 1$, 2, 3 and 4. All these results say that the sulfur alloying in $Cu_8SiS_xSe_{6-x}$ can be used to shift the polymorphic phase transition below room temperature and stabilize the high-temperature γ-phase over the broad temperature range.

The microstructure of the samples after SPS treatment was analyzed using optical microscopy under polarized light and scanning electron microscopy (Figures 3 and S3). The microstructure images of $Cu_8SiS_xSe_{6-x}$ bulk samples indicate good densification after the SPS procedure in agreement with the high density of samples (around 96 % of the theoretical density) measured by the Archimedes method. More details about the microstructure were highlighted by analyzing the optical microscopy images under polarized light. Particularly, the distribution of the grains in $Cu_{8-\delta}SiS_xSe_{6-x}$ samples with $x = 0$, $\delta = 0$ (Figure 3a), $x = 3$, $\delta = 0$ (Figure 3c), $x = 3$, $\delta = 0.2$ (Figure 3e), and $x = 6$, $\delta = 0$ (Figure 3g) looks different. This observation is especially interesting, as all powders before SPS were sieved using the same seave with a grain size of 50 μm, as well as the other technological procedures, were the same for all investigated samples. The grain size of the samples varies significantly between 10 μm and 70 μm (indicating the growth of grains during sintering). Moreover, the $Cu_{8-\delta}SiS_xSe_{6-x}$ samples with $x = 3$ show larger grains with good inter-grain contacts (Figure 3 c and e), while samples with $x = 0$ and $x = 6$ contain broad and smeared inter-grain regions. This observation can be connected with that, the $Cu_{8-\delta}SiS_xSe_{6-x}$ samples with $x = 0$ and $x = 6$ are crossing the phase transition α - γ during the SPS procedure, while the samples with $x = 3$ have existed as the γ-phase over the temperature range from room temperature up to sintering temperature during SPS (1023 K).

The chemical composition of the investigated $Cu_{8-\delta}SiS_xSe_{6-x}$ materials was determined by WDS collecting and averaging the chemical composition of ten regions marked by red points in the right panels in Figure 3. The WDS chemical composition of the investigated samples is in reasonable agreement with the nominal chemical composition, while a slight deviation of Cu still can be observed in samples with $x = 0$ and $x = 3$. The presence of Cu-based binary phases with a chemical composition close to the $Cu_{1.8}$(S,Se) phase was also registered on the grain boundaries, which agrees with the results of the XRD analysis. The EDS elemental mapping of the $Cu_8SiS_xSe_{6-x}$ sample with $x = 3$ (Figure S3f) shows the homogenous distribution of the elements in the main phase with the trace presence of binary copper-based chalcogenides.

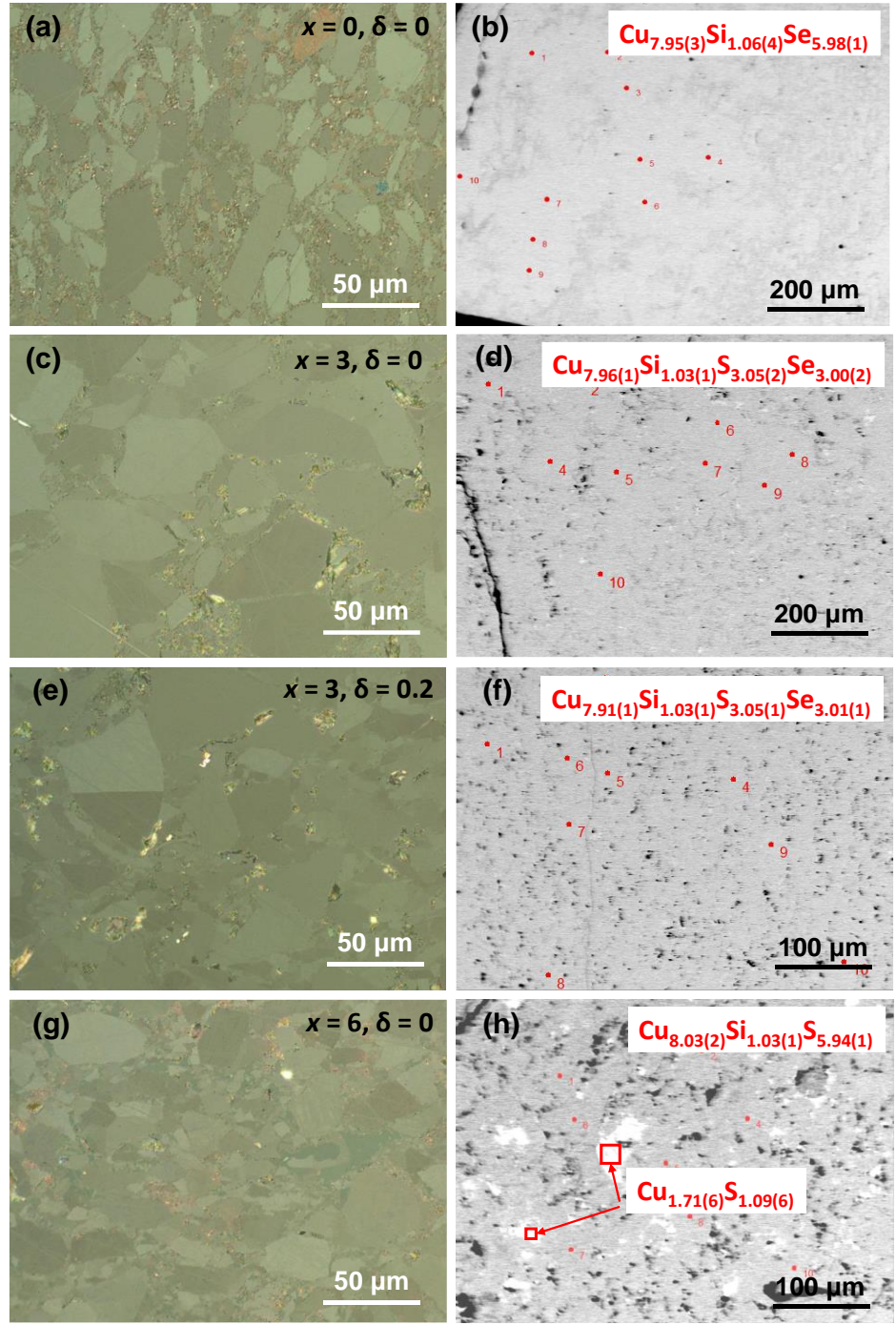

**Figure 3.** OM images under polarized light (left panels) and SEM images (right panels) of the $Cu_{8-\delta}SiS_xSe_{6-x}$ samples after SPS (a,b) $x = 0$, $\delta = 0$; (c,d) $x = 6$, $\delta = 0$; (e,f) $x = 3$, $\delta = 0$; (g,h) $x = 3$, $\delta = 0.2$. The chemical compositions of samples achieved using WDS analysis are marked in the SEM images.

## 3.3. Transport properties

### 3.3.1. Seebeck coefficient and electrical resistivity

The Seebeck coefficient and electrical resistivity of the $Cu_8SiS_xSe_{6-x}$ argyrodites at 300 K show an increasing trend with the rise of sulfur content (Table 1). The deviation from the general tendencies of the electronic transport properties, i.e. for samples with $x = 1$ and $x = 5$, can be connected with the significant effect of both polymorphic modifications α and γ in these materials. Considering the performed phase equilibria analysis, the dominative effect of the disordered α phase is expected for materials with $x = 0$ and 6, while the transport properties of $Cu_8SiS_xSe_{6-x}$ materials with $x = 2$ and 3 should be mainly determined by ordered γ phase. In turn, the properties of the rest samples can be determined by both polymorphs. The values of the electronic transport coefficients α and ρ well correspond to the Hall concentration $n_H$ measured for the investigated materials. The Seebeck coefficient for selenide ($x = 0$) is significantly lower compared to the α value of the other samples, which is connected with low-symmetry α-modification that existed in this sample at low temperatures. The large value of the Seebeck coefficient for sulfide ($x = 6$) can be explained by the very low carrier concentration of $1.4 \times 10^{18}$ cm$^{-3}$ recorded in this sample. The Hall mobility of all samples is in the range of 0.3-13.0 cm$^2$V$^{-1}$s$^{-1}$, which agreed with the previously reported data for sulfur and selenium-contained Cu-based argyrodites. The sample with $x = 1$ shows higher Hall mobility compared to other specimens, due to lower carrier concentration.

**Table 1.** Seebeck coefficient α, electrical resistivity ρ, Hall concentration $n_H$, Hall carrier mobility $\mu_H$, and density of states effective mass $m^*$ for $Cu_8SiS_xSe_{6-x}$ specimens at 300 K

| $Cu_8SiS_xSe_{6-x}$ | α, μV K$^{-1}$ | ρ, mΩ cm | $n_H$, cm$^{-3}$ | $\mu_H$, cm$^2$ V$^{-1}$ s$^{-1}$ | $m^*/m_e$ |
|---|---|---|---|---|---|
| $x = 0$ | 76 | 26 | $5.2 \times 10^{19}$ | 4.7 | 0.23 |
| $x = 1$ | 208 | 32 | $1.5 \times 10^{19}$ | 13.0 | 0.81 |
| $x = 2$ | 162 | 18 | $1.2 \times 10^{20}$ | 3.0 | 1.91 |
| $x = 3$ | 205 | 24 | $8.1 \times 10^{19}$ | 3.2 | 2.4 |
| $x = 4$ | 272 | 64 | $3.1 \times 10^{19}$ | 3.1 | 2.39 |
| $x = 5$ | 325 | $1.32 \times 10^3$ | $1.9 \times 10^{19}$ | 2.5 | 0.59 |
| $x = 6$ | 325 | $13.2 \times 10^3$ | $1.4 \times 10^{18}$ | 0.3 | 0.47 |

To further investigate the effect of the Se/S substitution on the electronic transport performance of the investigated argyrodites, the density of states effective masses $m^*$ were estimated. The calculations were performed employing the Kane band model via the procedure described in Supporting information and our previous papers [17,38]. Interestingly, the $m^*$ values sharply increase from $0.23m_e$ for the selenide to $2.4m_e$ for the $Cu_8SiS_xSe_{6-x}$ sample with $x = 3$ and then decrease to $0.47m_e$ for sulfide. In turn, the values of the carrier mobility do not follow this change.

Considering the scattering at the non-polar phonons (either acoustic or optical) as the main scattering mechanism (which has been found in most known and good thermoelectrics), the carrier mobility $\mu_H$ is inversely proportional to the value of the density of electronic states effective mass at carrier pocket $m_b^*$ [13,38,39]. Hence the dramatic change of the DOS effective mass $m^*$ observed for the investigated argyrodites can only be connected with the higher band degeneracy $N_V$ of the high-symmetry γ-modification. In turn, the α-modification with $N_v = 1$ (bandgap in $\Gamma$ point of the Brillouin zone) may produce only limited DOS effective masses. Large values of $m^*$ accompanied by the maintained mobility $\mu$ predict the enhancement of electronic transport properties [16,32,40].

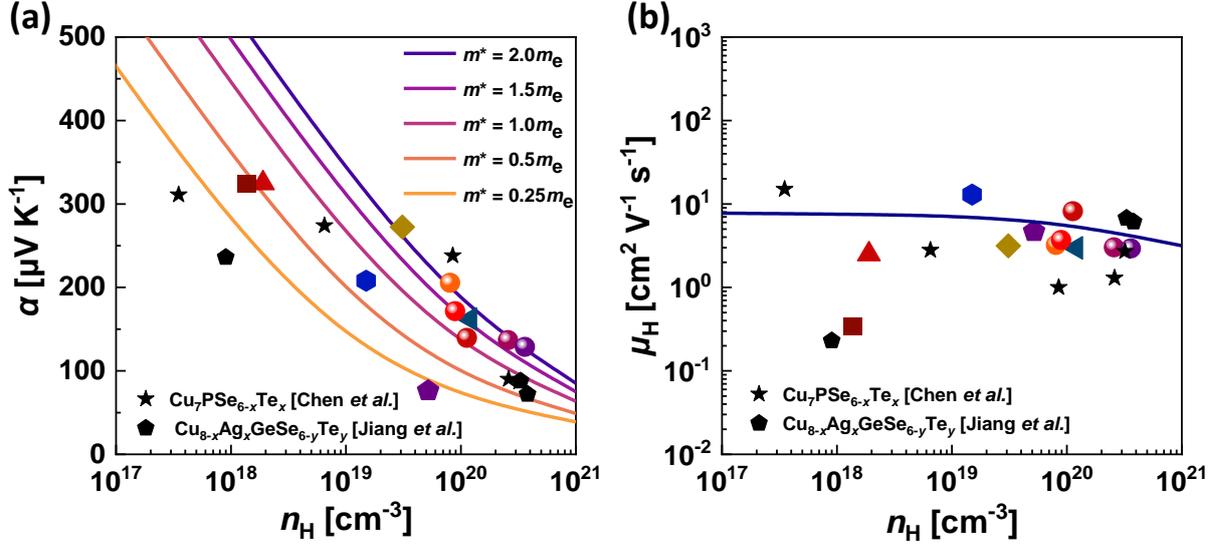

**Figure 4.** The concentration-dependent (a) Seebeck coefficient and (b) Hall mobility for $Cu_8SiS_xSe_{6-x}$ argyrodites at 300 K. Points correspond to the experimental measurements (black stars and pentagons have been taken from refs. [30,41], and other symbols correspond to the samples investigated in this work). Lines were calculated using the Kane band model considering acoustic scattering as the main scattering mechanism. The mobility calculations were performed using the conductivity effective mass of $2.0m_e$ and deformation potential of 10 eV.

The room temperature Seebeck coefficient and Hall mobility as a function of carrier concentration for the investigated tetrahedrites are plotted in Figure 4. We also employ the solution of the Boltzmann transport equation within constant scattering time approximation to estimate the theoretical dependences of the Seebeck coefficient and carrier mobility. All details of the calculations can be found in the Supporting information and our previous papers [9,17]. As it was mentioned above, the almost one-order change of the DOS effective mass was necessary to fit the experimental points with the theoretical Seebeck calculations. In turn, the data reported by Chen for $Cu_7PSe_{6-x}Te_x$ [30] and Jiang et al. for $Cu_{8-x}Ag_xGeSe_{6-y}$ [41] are also in good agreement with the obtained in our work results. Considering the above analysis of the $\alpha(n_H)$ dependence shown in Figure 4a, we can say that the low-symmetry phases show very low DOS effective masses, while the high-symmetry ones require much higher $m^*$ values. The mobility of the investigated argyrodites is rather low and shows only a minor decreasing trend over the broad concentration range. The experimental points from Refs. [30,41] and the obtained in this work data roughly correspond to the Kane band model theoretical prediction considering the value of the effective mass of $2.0m_e$.

The Seebeck coefficient as a function of temperature for $Cu_8SiS_xSe_{6-x}$ samples is shown in Figure 5a. All samples possess a positive Seebeck coefficient $\alpha$ over the entire temperature range, indicating that holes are the dominant charge carriers. Seebeck coefficient for the pristine $Cu_8SiSe_6$ and $Cu_8SiS_6$ shows the lowest and the highest trends, respectively, while the investigated S/Se substituted solid solutions are in between. Such a tendency can be well explained considering the different concentrations as well as the different bandgap values of and for selenide ($E_g = 1.33$ eV) and sulfide ($E_g = 1.84$ eV), respectively [42]. The Seebeck coefficient for pristine $Cu_8SiSe_6$ shows the increasing tendency over the investigated temperature range, and with the addition of S becomes more temperature independent. Also, it should be mentioned, the fluctuation of the Seebeck coefficient at low temperatures for the endmember samples can be explained by the polymorphic phase transition from the α- to γ-modification. Moreover, we should keep in mind that the presence of both modifications can have an effect on the values of the Seebeck coefficient at the whole investigated temperature range.

Figure 5b shows the electrical resistivity for $Cu_8SiS_xSe_{6-x}$ materials over the entire temperature range of 298-773 K. The value of $\rho$ for the investigated $Cu_8SiS_xSe_{6-x}$ materials decreases with increasing temperature, indicating a semiconductor behavior. In $\alpha(T)$ and $\rho(T)$ trends, we do not observe typical for the narrow bandgap semiconductors minority carrier transport, probably due to relatively wide bandgaps. In agreement with the $\alpha(T)$ dependencies, the electrical resistivities for pristine $Cu_8SiS_6$ and $Cu_8SiSe_6$ materials show the highest and the lowest trends, respectively. The lowest trend of $\rho$ is observed for the samples with $x = 1 - 3$, which is connected with only γ-modification observed in the XRD pattern for these samples. Also, the sharp decreasing tendency of $\rho(T)$ was observed for sulfide and the sample with $x = 5$, while with increasing Se content, the $\rho(T)$ dependence tends to be almost temperature independent. This observation can indicate the change of the dominant charge scattering mechanism for $Cu_8SiS_xSe_{6-x}$ samples with S/Se substitution.

To evaluate the potential of the developed materials and test the performance of the S/Se substitution for tuning the electronic transport, we have calculated the weighted mobility for the investigated $Cu_8SiS_xSe_{6-x}$ materials [10]. Interestingly, the temperature-dependent weighted mobility for $Cu_8SiS_xSe_{6-x}$ samples with $x = 0, 5$, and 6 shows increasing trends, while $\mu_w(T)$ for the rest samples decreases with temperature (Figure 5c). This observation indicates the change of the dominant carrier scattering mechanism from defect scattering or/and ionized impurity scattering to acoustic phonon scattering [10]. It should also be mentioned that the samples with the decreasing $\mu_w(T)$ trends show much higher weighted mobility compared to the other ones, indicating their high potential for further optimization through carrier concentration tuning.

We also evaluate the configurational entropy change ($\Delta S = -R \sum_{i=1}^{n} x_i \ln(x_i)$, where $R$ is the universal gas constant and $x_i$ is the composition of each sample [29]) which was achieved due to the Se/S elemental substitution in $Cu_8SiS_xSe_{6-x}$. Although the investigated samples are in the range of the low-entropy alloys, the increase of entropy up to $\Delta S = 0.4$-$0.7$ R was sufficiently high to decrease the polymorphic phase transition of the investigated argyrodites and receive the high-symmetry γ-phase over the broad temperature range of 300-773 K.

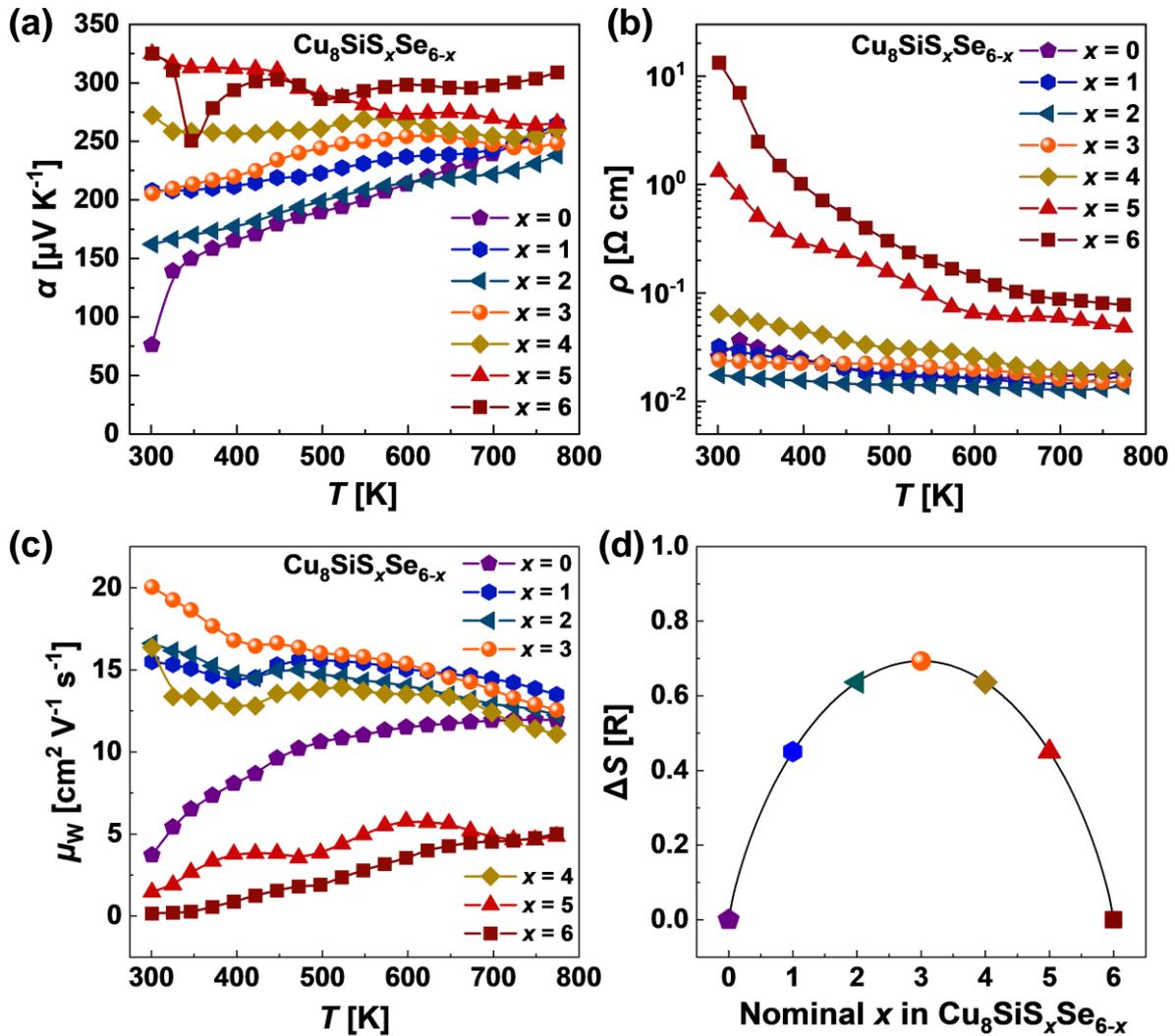

**Figure 5.** Temperature-dependent (a) Seebeck coefficient, (b) electrical resistivity (c) weighted mobility, and (d) composition-dependent configurational entropy for $Cu_8SiS_xSe_{6-x}$ materials.

The plot of the $\ln[\rho]$ on the $1/T$ and the estimated activation energies for $Cu_8SiS_xSe_{6-x}$ is shown in Figure S5 (Supporting information). The activation energy increase with the $x$ increase, following the expected tendency of the bandgap, which should continuously decrease from sulfide to selenide. The exception here is the $Cu_8SiS_xSe_{6-x}$ sample with $x = 2$, which shows somewhat lower activation energies considering the continuous increasing trend. This can be attributed to the significantly higher carrier concentration observed in this sample (Table 1).

### 3.3.2. Thermal conductivity and thermoelectric performance

The temperature-dependent total thermal conductivity for $Cu_8SiS_xSe_{6-x}$ alloys is shown in Figure 6a. All samples possess extremely low thermal conductivities, in the range of ~ 0.25 - 0.35 W

m$^{-1}$ K$^{-1}$ over the entire temperature range, which is among the lowest values observed in crystalline materials.

To analyze the phonon transport, we estimated the lattice thermal conductivity $\kappa_L$ subtracting the electronic contribution $\kappa_e$ from the total thermal conductivity $\kappa$. The values of $\kappa_e$ were calculated using the Wiedemann-Franz law $\frac{\kappa_e}{\sigma} = LT$, where $L$ is the Lorenz number. The calculations were performed using the Kane band model employing the acoustic phonon scattering approximation and effective masses obtained by experimentally measured carrier concentration and Seebeck coefficients. The details of the calculations are shown in the Supporting information. Although the used approximation can give only a rough estimation of the $\kappa_e$, we believe that it is acceptable for the case of the Cu$_8$SiS$_x$Se$_{6-x}$ argyrodites due to relatively high values of electrical resistivity and a generally small amount of heat transported by electrons.

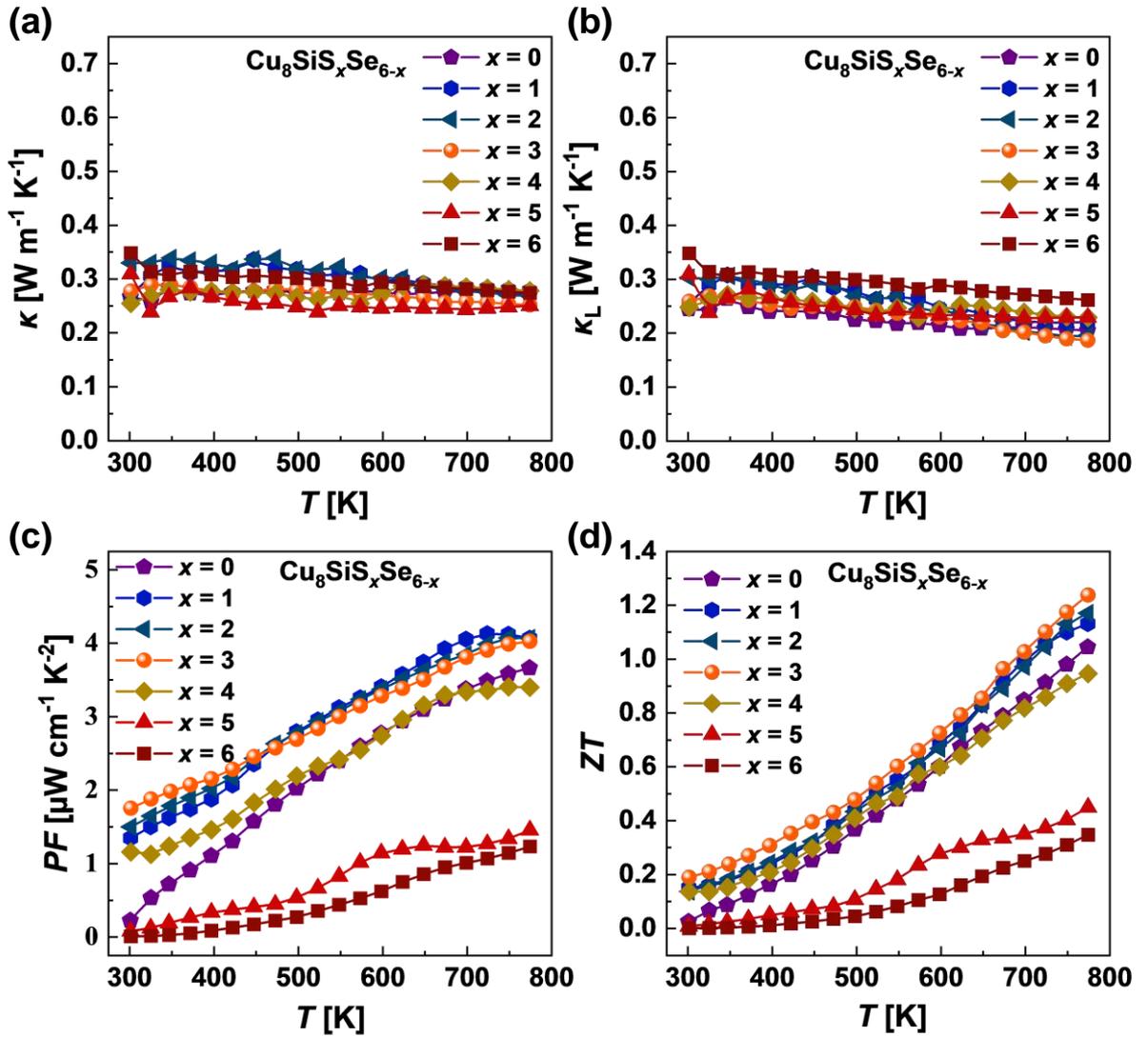

**Figure 6.** Temperature-dependent (a) total and (b) lattice thermal conductivity, (c) power factor, and (d) dimensionless thermoelectric figure of merit *ZT* for the Cu$_8$SiS$_x$Se$_{6-x}$ materials.

Figure 6b shows the lattice thermal conductivity as a function of temperature for $Cu_8SiS_xSe_{6-x}$ argyrodites. It can be found, that the $\kappa_L(T)$ ranges from 0.19 Wm$^{-1}$K$^{-1}$ to 0.35 Wm$^{-1}$K$^{-1}$ with a weak decreasing tendency over the investigated temperature range. Such ultralow values of the $\kappa_L$ are significantly lower compared to the most state-of-the-art TE materials [25,43–45]. In turn, most of the known argyrodites show the ultralow lattice thermal conductivity over the broad temperature range, which is explained in the literature through the migration or "liquid-like behavior" of Cu-ions through the rigid framework built with the chalcogen atoms. The temperature trend of the lattice thermal conductivity for selenide is the lowest while for sulfide $\kappa_L$ is the highest. The exception here is the temperature trends for the $Cu_8SiS_xSe_{6-x}$ materials with $x = 2$ and 3, which are even lower than for the other materials at higher temperatures probably due to the highest point defect scattering. This observation is also agreed with the highest Hall carrier concentration measured for these samples (Table 1).

The ultrasonic measurements performed for the investigated argyrodites are shown in Table S1 (Supporting information). The average speed of sound $v_m$ for $Cu_8SiS_xSe_{6-x}$ is in the range of 2107-2362 ms$^{-1}$, which is similar or even higher compared to the previously reported Cu-based argyrodites (1700-2100 ms$^{-1}$) [32,41,46]. The obtained values of the $v_m$ are also higher compared to the reported speed of sound for Ag-based argyrodites (1350-2000 ms$^{-1}$) [1,47–50]. While the relatively low values of the speed of sound in Ag-based argyrodites potentially can be responsible for the ultralow lattice thermal conductivity, the obtained in this work and reported relatively high values of $v_m$ for Cu-based argyrodites is contradicting the classical explanation of the phonon transport in solids. Particularly, the speed of sound in solids is closely related to the speed of sound through the relation $\kappa_L = 1/3 C_v v l$, where $C_v$ is the heat capacity and $l$ is the phonon mean free path, hence the high speed of sound in Cu-based argyrodites with ultralow lattice thermal conductivity seems to be intriguing. A good explanation of this effect was proposed by Weldert et al. [46]. The authors, explaining the ultralow $\kappa_L$ in $Cu_7PSe_6$ argyrodite, assumed that the large portion of the phonon modes does not propagate through the lattice at all. Following this assumption and considering the huge difference in the longitudinal and transverse speeds of sound, we also hypothesize the softening of the shear phonon modes in the investigated materials due to the liquid-like Cu migration as the main reason for the ultralow $\kappa_L$.

The ultrasonic measurements also allowed us to estimate the Grüneisen parameters $\gamma$ for the investigated argyrodites, which ranged from 1.75 to 1.97 (Table S1). Such high values of the $\gamma$ were also reported for the other Cu- and Ag-based argyrodites [1,32,41,47–50]. The high Grüneisen parameters suggest the strong anharmonicity of lattice vibrations, which can be the other reason for the ultralow lattice thermal conductivity observed in the investigated $Cu_8SiS_xSe_{6-x}$ argyrodites.

To evaluate the energy conversion performance of the developed $Cu_8SiS_xSe_{6-x}$ argyrodites, we have calculated the power factors ($PF = \alpha^2/\rho$) and dimensionless thermoelectric figure of merit $ZT$ (Figures 6c and d). $Cu_8SiS_6$ has the lowest power factor, while the *PF* for $Cu_8SiSe_6$ is around three times higher. In turn, the highest power factors over the entire temperature range are observed for the samples with $x = 1-3$, which is connected with the optimized Seebeck coefficient [51] and reduced electrical resistivity. Also, we should keep in mind that these samples are characterized by the domination of the high-symmetry γ-phase. As a result of the enhanced *PF* and ultralow thermal conductivity, a significant improvement of the dimensionless thermoelectric figure of merit *ZT* has been obtained for $Cu_8SiS_xSe_{6-x}$ solid solutions compared to the pristine $Cu_8SiS_6$ and $Cu_8SiSe_6$ samples. A maximum *ZT* value of ~ 1.2 is achieved for the $Cu_8SiS_xSe_{6-x}$ sample with $x = 3$ at 773 K, opening the great potential of the developed materials for TE applications.

## 4. Enhancement of the electronic transport using Cu deviation

Although the *ZT* values for the $Cu_8SiS_xSe_{6-x}$ samples with higher entropy are quite similar over the investigated temperature range, it can be observed that the best temperature trends of the power factor correspond to the highest carrier concentrations (Table 1) indicating unoptimized carrier concentration [52]. Consequently, the further increase of the carrier concentration may result in even improved *PF*. As it is mentioned above and discussed in the literature, the carrier concentration of argyrodites is governed by the Cu vacancies, hence effective chemical potential tuning is expected with the deviation of Cu stoichiometry. To further improve the TE performance of the investigated argyrodites, we prepared an additional series of samples with the chemical composition of $Cu_{8-\delta}SiS_3Se_3$ ($\delta$ = 0.025, 0.05, 0.01, 0.02).

The powder XRD patterns of $Cu_{8-\delta}SiS_3Se_3$ samples after synthesis are shown in Figure S4a (Supporting information). All registered XRD patterns were indexed in the cubic *F*-43*m* space group, which corresponds to the high-temperature γ-polymorph. Although the tiny impurity peaks of $Cu_2(S,Se)$ at around 27° and 45° were noticed, their intensity was decreasing with increasing $\delta$ in $Cu_{8-\delta}SiS_3Se_3$, which can indicate the deviation of stoichiometry from the nominal $Cu_8SiS_3Se_3$. In turn, the increase of Cu-deficiency in $Cu_{8-\delta}SiS_3Se_3$ leads to a roughly linear decrease of lattice parameters (Figure S4b, Supporting information), suggesting a shrinkage of the lattice without any noticeable change in the crystal structure symmetry.

**Table 2.** Seebeck coefficient $\alpha$, electrical resistivity $\rho$, Hall concentration $n_H$, Hall carrier mobility $\mu_H$, and density of states effective mass $m^*$ for $Cu_{8-\delta}SiS_3Se_3$ specimens at 300 K

| $Cu_{8-\delta}SiS_3Se_3$ | $\alpha$, µV K$^{-1}$ | $\rho$, mΩ cm | $n_H$, cm$^{-3}$ | $\mu_H$, cm$^2$ V$^{-1}$ s$^{-1}$ | $m^*/m_e$ |
|---|---|---|---|---|---|
| $\delta = 0$ | 205 | 24 | $8.1\times10^{19}$ | 3.2 | 2.4 |
| $\delta = 0.025$ | 172 | 19 | $8.9\times10^{19}$ | 3.7 | 1.75 |
| $\delta = 0.05$ | 140 | 6.8 | $1.1\times10^{20}$ | 8.2 | 1.35 |
| $\delta = 0.1$ | 137 | 8.0 | $2.6\times10^{20}$ | 3.0 | 2.23 |
| $\delta = 0.2$ | 129 | 5.9 | $3.6\times10^{20}$ | 2.9 | 2.45 |

Table 2 shows the electronic transport properties in Cu-deficient $Cu_{8-\delta}SiS_3Se_3$ samples. It can be observed, that the carrier concentration increases significantly from the $8.1\times10^{19}$ cm$^{-3}$ up to $3.6\times10^{20}$ cm$^{-3}$ with the increase of copper deficiency. The increase in $n_H$ results in a decrease of the Seebeck coefficient $\alpha$ and electrical resistivity $\rho$ following the classical electronic transport behavior in semiconductor materials. Although we recorded a rather weak dependence between the carrier concentration and Hall mobility (Table 2), a large change in the density of state effective mass was found in this series. Particularly, the DOS effective mass is the highest for the sample without Cu deviation ($\delta = 0$) and for the sample with the highest investigated Cu deviation ($\delta = 0.2$), while the lowest effective mass of $1.35m_e$ is observed for the $Cu_{8-\delta}SiS_3Se_3$ sample with $\delta = 0.05$ (Table 2). But what is even more interesting, in contradiction to the first series of samples, the lowest DOS effective mass $m^*$ is observed simultaneously with the highest value of carrier mobility $\mu_H$. Hence, the Cu-deviation rather does not change the number of carrier pocket $N_v$ (like in the case of S substitution in $Cu_8SiS_xSe_{6-x}$) but only effectively modifies the carrier density $n_H$.

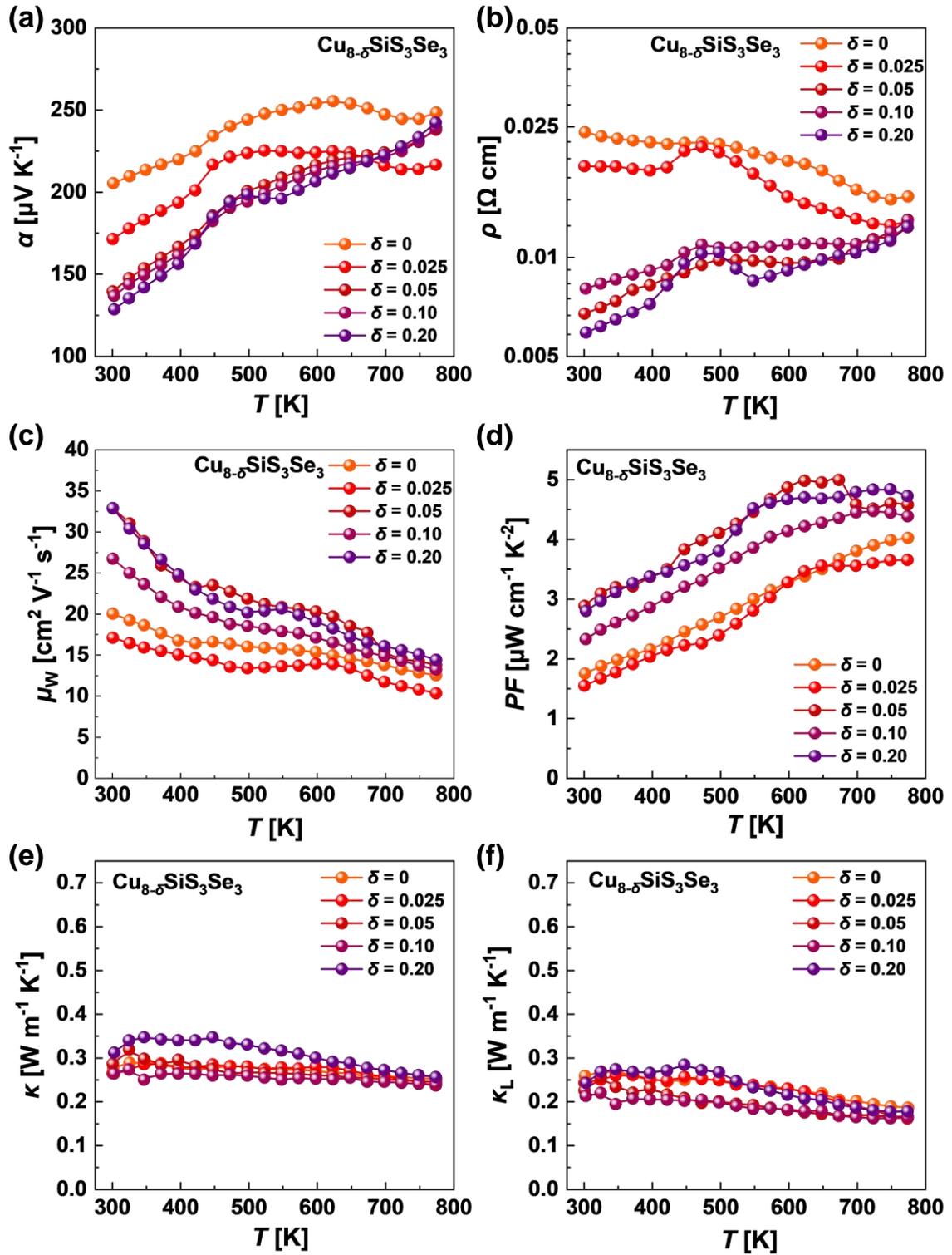

**Figure 7.** Temperature-dependent (a) Seebeck coefficient, (b) electrical resistivity, (c) weighted mobility (d) power factor, (e) total, and (f) lattice thermal conductivity for $Cu_{8-\delta}SiS_3Se_3$ materials.

The temperature-dependent thermoelectric properties of the $Cu_{8-\delta}SiS_3Se_3$ argyrodites are shown in Figure 7. Due to the increase in carrier concentration, the copper deficiency results in a decrease in the Seebeck coefficient and electrical resistivity (Figures 7a and b). The temperature trends of the electrical resistivity $\rho(T)$ also change the slope from negative to positive indicating the change of the carrier transport from semiconductor to metal-like one (Figure 7b). As a result of the enhanced carrier concentration, the weighted mobility and power factor show higher values for the samples with higher Cu deficiency (Figures 7c and d). However, the best weighted mobility and power factor is belonging to the $Cu_{8-\delta}SiS_3Se_3$ sample with δ = 0.5, due to the lower effective mass and higher carrier mobility observed in this material.

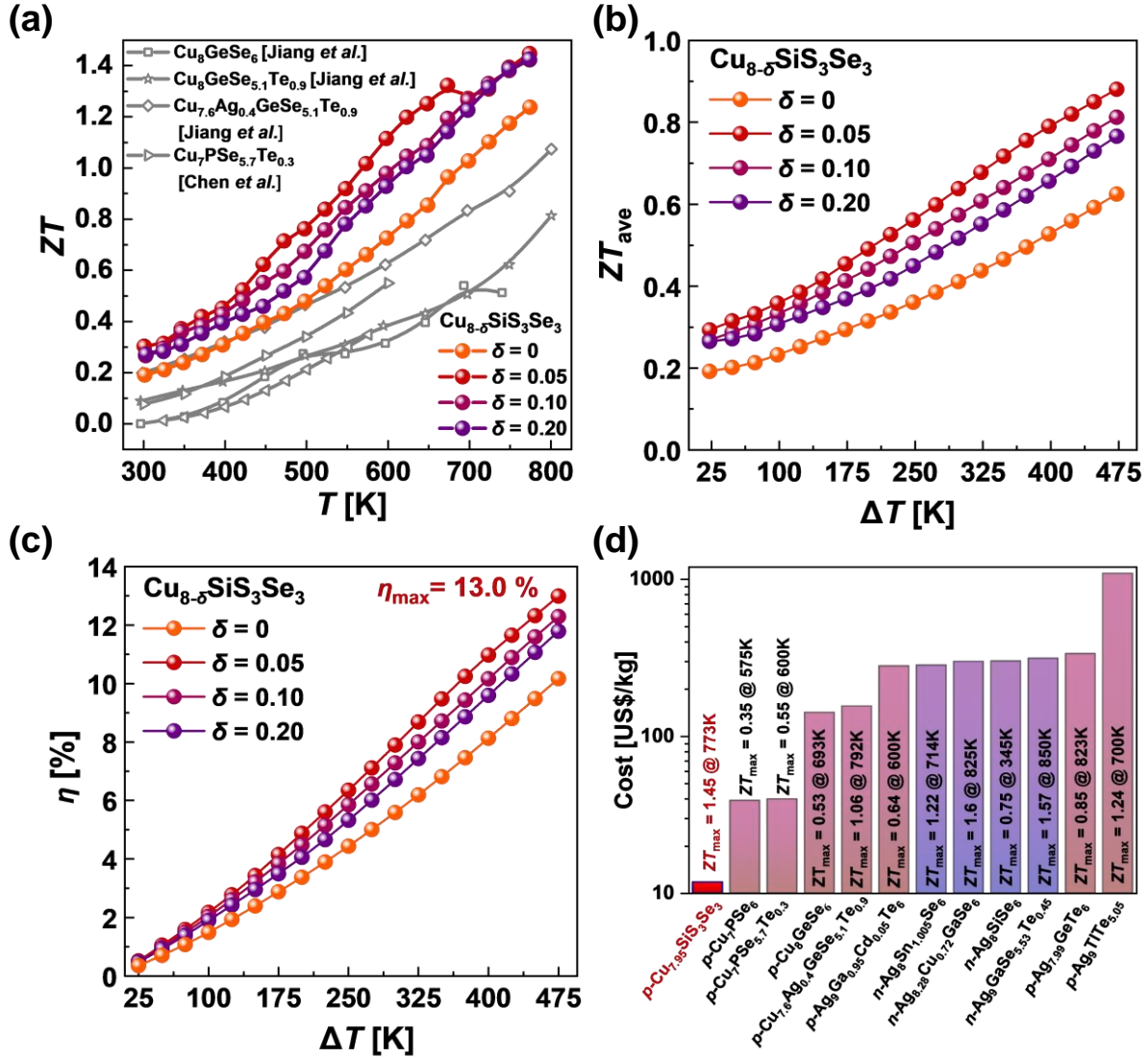

**Figure 8.** (a) Comparison of the obtained figure of merit ZT values with those reported in the literature for Cu-based argyrodites [30,41]; (b) averaged figure of merit $ZT_{ave}$ and (c) estimated efficiency for $Cu_8SiS_xSe_{6-x}$ (in calculating $ZT_{ave}$ and η, we assumed the cold temperature $T_c$ = 300 K); (d) summary of cost (in US$/kg) of the best n-type [48,53–55] and p-type [30,41,46,56–58] argyrodite thermoelectric materials (price source: Shanghai metals market, SMM).

The enhanced electronic transport together with the well-maintained ultralow thermal conductivity (Figures 7e and f) leads to significant improvement of the ZT parameter and thermoelectric quality factor B (Figures 8a and S6). Particularly, the dimensionless thermoelectric figure of merit ZT and parameter B achieve a very high value of 1.45 and 0.6 at 773 K for $Cu_{8-\delta}SiS_3Se_3$ samples with δ = 0.05-0.2, respectively. Although the ZT parameter is the highest reported up-to-date for *p*-type argyrodites, the real practical interest of the material should be evaluated using the average thermoelectric figure of merit $ZT_{ave}$ ($ZT_{ave} = \frac{1}{T_h - T_c} \int_{T_c}^{T_h} ZT \cdot dT$, where $T_h$, $T_c$ denote the hot side and cold side temperatures, respectively). Hence we performed this estimation and found that the $ZT_{ave}$ for the best material $Cu_{7.95}SiS_3Se_3$ achieved 0.9 at $\Delta T = 475$ K ($\Delta T = T_h$-$T_c$) (Figure 8b), which is the highest value any time obtained for the Cu-based argyrodites. The estimated maximum energy conversion efficiency $\eta_{max}$ ($\eta_{max} = \frac{\Delta T}{T_h} \frac{\sqrt{1+ZT_{ave}} - 1}{\sqrt{1+ZT_{ave}} + \frac{T_c}{T_h}}$, where $\Delta T$ is the temperature difference between the hot and cold sides ($\Delta T = T_h - T_c$)) for the TE leg developed using the $Cu_{8-\delta}SiS_3Se_3$ reach 13.0 % (Figure 8c). Furthermore, the utilization of earth-abundant and low-cost elements dramatically lowers the costs of the developed material to the outstanding level of ~ 12 US$/kg (Figure 8d) offering the developed material for energy conversion applications.

## 4. CONCLUSIONS

This work discovers the great thermoelectric performance in new $Cu_8SiS_xSe_{6-x}$ argyrodites. To achieve the stable *γ*-modification over the application temperature range we employ the entropy engineering approach. Particularly, we found that the S/Se substitution produces the sufficient increase in configurational entropy necessary for shifting the polymorphic phase transition (from *α*- to *γ*-modification) below the room temperature. The preferable *γ*-phase enables the high values of the DOS effective masses without degradation of the carrier mobility. As a result, a high Seebeck coefficient and enhanced weighted mobility are observed over the broad temperature range of 298 - 773 K. The lattice thermal conductivity for all investigated samples is in the range of 0.19 - 0.35 Wm$^{-1}$K$^{-1}$, which is among the lowest values of $\kappa_L$ reported for the crystalline materials up to date. As a result of the remarkably enhanced electronic transport and ultralow lattice thermal conductivity, the TE figure of merit *ZT* reaches a very high value of 1.2 at 773 K. The further tuning of the carrier concentration through the Cu deficiency in $Cu_{8-\delta}Si_8S_3Se_3$ argyrodite materials increases the maximum *ZT* up to 1.45 at 773 K. The estimated energy conversion efficiency achieves the outstanding value of 13.0 % at a temperature gradient of 475 K ($T_c$=300 K) for $Cu_{8-\delta}SiS_3Se_3$ samples with $\delta = 0.05$ due to the enhancement of the average $ZT_{ave}$. The calculated price of 12 US$/kg and the great thermoelectric performance make the discovered $Cu_{7.95}SiS_3Se_3$ argyrodite a real candidate for energy conversion applications.

**ACKNOWLEDGMENTS**

The research was funded by the Foundation for Polish Science (TEAM-TECH/2016-2/14 Grant "New approach for the development of efficient materials for direct conversion of heat into electricity"), co-financed by the European Union under the European Regional Development Fund. T.P. acknowledges support from program „Excellence Initiative – Research University" for the AGH University of Science and Technology.